# Neural Network Revisited: Perception on Modified Poincare Map of Financial Time Series Data[1]

**Hokky Situngkir**
(hokky@elka.ee.itb.ac.id)
Dept. Computational Sociology
Bandung Fe Institute

**Yohanes Surya**
(yohaness@centrin.net.id)
Dept. Physics
Universitas Pelita Harapan

## Abstract

Artificial Neural Network Model for prediction of time-series data is revisited on analysis of the Indonesian stock-exchange data. We introduce the use of Multi-Layer Perceptron to percept the modified Poincare-map of the given financial time-series data. The modified Poincare-map is believed to become the pattern of the data that transforms the data in time-t versus the data in time-t+1 graphically. We built the Multi-Layer Perceptron to percept and demonstrate predicting the data on specific stock-exchange in Indonesia.

There has been a lot of works on forecasting and prediction of the future value using neural network perception, e.g.: the previous paper, Situngkir & Surya (2003) that explained how to utilize multi-layer perceptron in predicting time series data. We actually can obtain, using this analysis, sufficiently good prediction of two up to three steps of time forward. It should be admitted that for the sake of prediction, the very wish of a predictor is to predict time series data in a wide length and as precise as possible. We also show that basically prediction analysis done was a *top*-down prediction analysis, means viewing a problem from such time series data, without concerning aspects caused the time series data.

Poincare map is simply a map that showing a pattern from its time series data. Poincare map is a map that is not time series map, yet it allows transversal changes of time series data in each time of iteration; in this way each element of data displayed can no longer be viewed differently from time series that represent them. In fact, every time series data has been inherent with the data that graphically represented. It should be admitted that Poincare map has been rarely used in social analysis. Traditional approaches identifying chaos (such as Lyapunov, correlation dimension and or spectral analysis) are rarely used in analyzing social system. This is factual since those ways are not sensitive to the poor amount of data (Marion, et al., 1997). We know that in the social system, data gained, are usually few, except financial time series-something that we would study further in the rest of this paper.

---

[1] Draft version of the paper to be presented in the Applications of Physics in Financial Analysis 4 (APFA4) Europhysics Conference of European Physical Society - Statistical and Nonlinear Physics Division, Warsaw, 13-15 November 2003, Warsaw University of Technology.

We will discuss how neural network – in this case the multi-layer perceptron - percepts a time series data that is previously transformed into modified Poincare map. This is suitable for a large number of data (up to thousands order) from the data we will predict. This paper is divided into three sections. First, we will elaborate what and how Poincare map is and where is its strong point. In the second section, the writer will show how neural network be used in percept Poincare map and what benefit we will get using this map perception. The third section will give picture of some implementation of Poincare map in several instances. As an extended part of the previous paper (Situngkir & Surya, 2003), we will use a long time series data of the fluctuating close stock price of PT TELKOM Indonesia. This will be followed by couples of conclusion notes and further possible development.

## 1. Poincare Map as A Pattern Map

Most physicists observe dynamical system by using calculus and differential equation to describe the whole movement of a system. Henri Poincare, a French mathematician once offered a step of work that simplifies a tiring time series data processing process. Poincare map can be described as a process in which every dynamic trajectory placed on a sheet of paper. On that paper, there would appear another picture of trajectory orbit from those time series data. This is illustrated in figure 1.

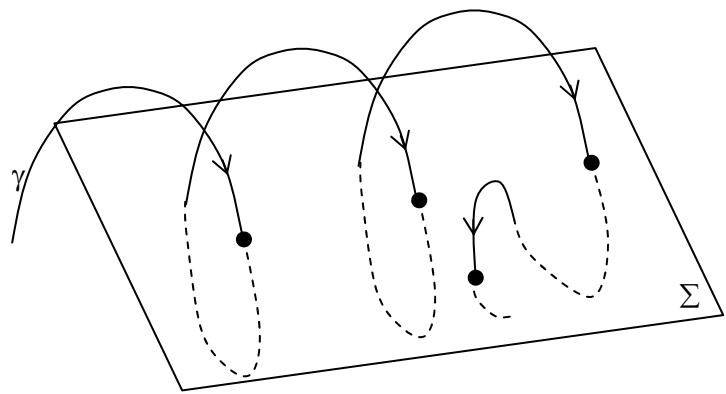

**Figure 1**
Sketch of Poincare Map as a dot pattern on a paper placed in trajectory
orbit of universal data

From this Poincare map basically we can see few concerns that for periodic systems, number of dots in Poincare map is limited on certain structure and on each dot allows same repeating. In the other hand, number of dots will be seen arbitrarily many and there will be no repeating dots for a-periodic system. Furthermore, in a *random* dynamical system Poincare map will exhibit without structure, and some dots may repeat. Basically, mathematics described by Poincare map are quite simple to solve than differential equation for dynamic trajectory data (Deane, et al., 1998).

Mathematically, we can say in short that Poincare map is a study of a transversal cut-plane $\Sigma$ from an $\gamma$ orbit trajectory with the flow of $\phi(t)$ where

$$x' = f(x), x \in R^n \tag{1.1}$$

It is obvious that to analyze $\gamma$ mapping to $\Sigma$ is easier than analyzing trajectory of dynamic system with full-n-dimensions. In its implementation, we can say:

$$x' = f(x) : U \subset R^n \to R^n \tag{1.2}$$

Flow from γ represented by *ϕ(t)* is obtained from periodic function ***T***, thus:

$$\phi(t+T, x_0) = \phi(t, x_0) \tag{1.3}$$

In this mapping we can see a dimensional reduction happens. This is caused by mapping function represented by *γ* (as *ϕ(t)*) which is time-varying function t has been reduced by eliminating dimension ***t*** and by displaying it topographically onto a of *Σ* cut plane. Thus we can understand that analysis using this map is basically an endeavor to constructing a simpler analysis space. For detail explanation and proving some theorems concerning with Poincare map characteristics for periodic system can be seen in (Tu, 1994:183-185) and (Hale, et al., 1991:375-377), and of chaotic and adaptive system in (Schuster, 1995:14-17).

Beside the simplicity, a Poincare map in fact gives more than things that we can imagine in enriching our analysis of dynamical system. In the analysis we perform here, surely the stock data function cannot be regarded as a periodic data. In this way we use generator function ***T*** as day period. In other words, we apply some modifications onto financial Poincare map.

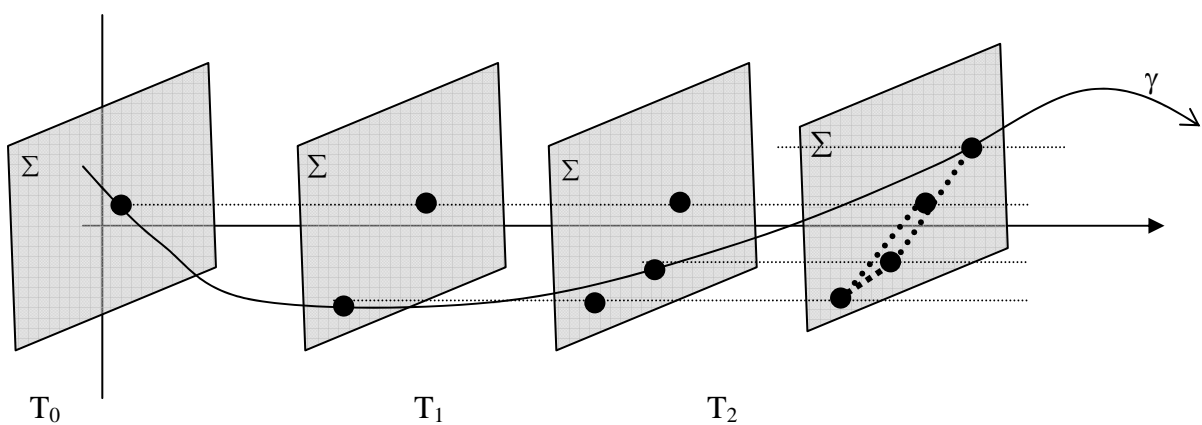

**Figure 2**
Modified Poincare map to predict financial data. We unchanged period to "picture"
financial data γ onto a cut plane *Σ*

Modification we are dealing about is that we decide generation function ***T*** by looking T as the period in which we "portrait" the movement of stock given (see figure 2). From here, we get cut plane, *Σ* as Poincare map that shows functional relation:

$$\phi(t, x_0) \in R^n \rightarrow \Sigma : \phi'(t, t+T) \in R^n \tag{1.4}$$

And the modified Poincare map represented by cut plane *Σ* becomes vector matrix that we will percept by the multi-layer perceptron for prediction purposes. In short, we transform the time series data into pattern data that is inherent with time data represented, and it is this pattern, that we try to analyze further with a hope that approximation and prediction using artificial neural network will become better.

## 2. The origins of patterns: correlations among data

As previously cited above, we try to analyze pattern that mapped in the financial Poincare map. It has been cited that artificial neural network perception on financial data transformed first into the Poincare map by drawing financial series data plot at time ***t***, that is ***y(t)*** mapped versus ***y(t+k)***. In this section we will find out frequency probability (***k*** value) that forms pattern of analyzed financial data. For this purpose, we find 'correlation' between one data and the other using autocorrelation principal: how one data correlates with its neighbors.

Autocorrelation principally is a stochastic processing data that is advantageous in getting partial description in cases of predicting time series data. This principal is very advantageous in detecting non-randomness of data and to identify which time series data model we are going to apply if our data is as a matter of fact not-random. Autocorrelation coefficient measures level of correlation among neighboring data in time series data.

As it explained in Nist-Sematech (2003), that if data stated in time series $y_i$ with $i = 1,2,3,...$, henceforth autocorrelation coefficient may be written as:

$$r_k = \frac{\sum_{i=1}^{n-k}(y_{i+k} - \overline{y}_{i+k})(y_i - \overline{y}_k)}{\sqrt{\sum_{i=1}^{n}(y_i - \overline{y}_i)^2 \times \sum_{i=1}^{n}(y_{i+k} - \overline{y}_{i+k})^2}} \qquad \ldots(2.1)$$

where $r_k$ is autocorrelation of $y_i$ and $y_{i+k}$. For some instances of data, autocorrelation coefficients form value distribution of about *k* that commonly called autocorrelation *sampling* distribution.

Here we can obtain autocorrelation plot that is formed by autocorrelation coefficient in auto-covariance function as the vertical axis and has always been in interval $[-1,1]$,

$$R_h = \frac{C_h}{C_o} \qquad \ldots(2.2)$$

with autocovariance function as

$$C_h = \frac{1}{N}\sum_{i=1}^{N-h}(y_i - \overline{y})(y_{i+h} - \overline{y}) \qquad \ldots(2.3)$$

and variance function as

$$C_o = \frac{\sum_{i=1}^{N}(y_i - \overline{y})^2}{N} \qquad \ldots(2.4)$$

and horizontal axis as time left $h = 1,2,3,...$.

Note that from equation (2.1) we may say that autocorrelation *sampling* distribution coefficient $r_k$ is normal distribution for

$$\mu_{r_k} = 0 \qquad \ldots(2.5)$$

and

$$\sigma_{r_k} = \frac{1}{\sqrt{n}} \qquad \ldots(2.6)$$

where $\mu_{r_k}$ and $\sigma_{r_k}$ respectively mean value and variance of $r_k$. To decide whether a data is a series of random or not random, then generally we can say that we use the boundary:

$$-2\sigma_{r_k} \leq r_k \leq 2\sigma_{r_k} \qquad \ldots(2.7)$$

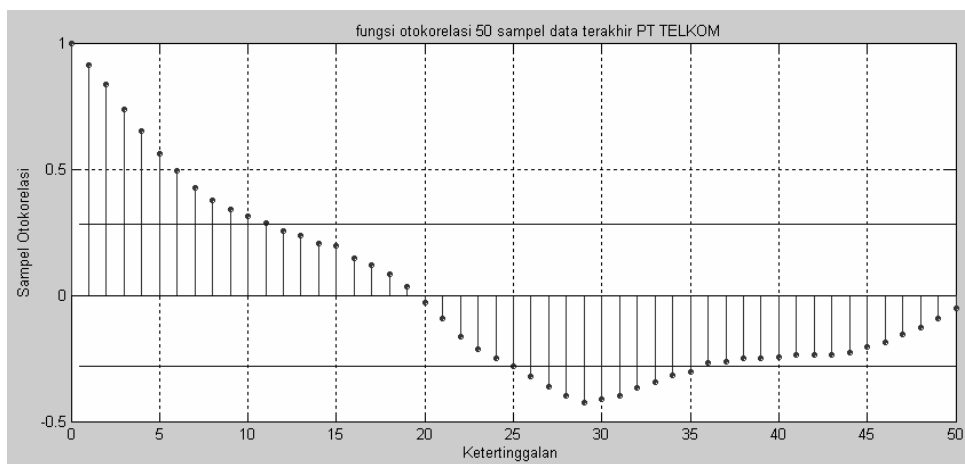

**Figure 3**
Autocorrelation plot on the last 50 data values of PT TELKOM

In Figure 3, we can see the autocorrelation plot of some financial time series data of stock market in Indonesia: PT TELKOM. On that graphic we can see clearly how the autocorrelation occurs between one data with the other ones. It can be seen that one data correlates in highest level with its previous and its next data, in so that data $y_i$ correlates in highest value with data $y_{i+1}$ and data $y_{i-1}$, which in consequence we have to choose **k=1** in perception of Poincare map we will perform.

## 2. Neural Network Perception in Poincare Map

Neural network model we use is a multi-layer perceptron as it has been used to percept time series data in Situngkir & Surya (2003). The different thing now is that we are no longer dividing data as we usually do for interval training, validation and examination. Using a diagram block we illustrate working process in Figure 4.

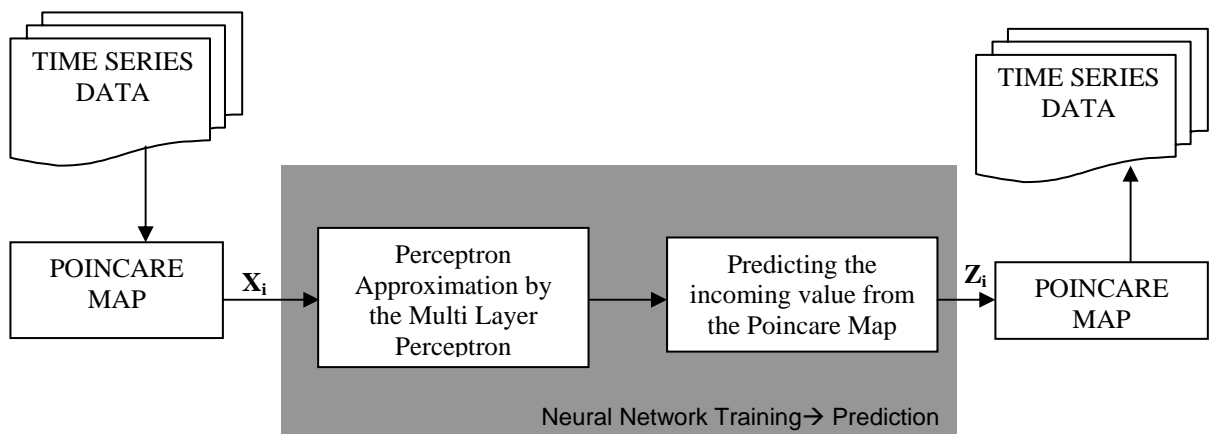

**Figure 4**
Working steps done in Neural Network Perception in Poincare Map

The type of the neural network training we use here is back-propagation training in regarding the error factor

$$E = \sum_i (Z_i - X_i)^2$$

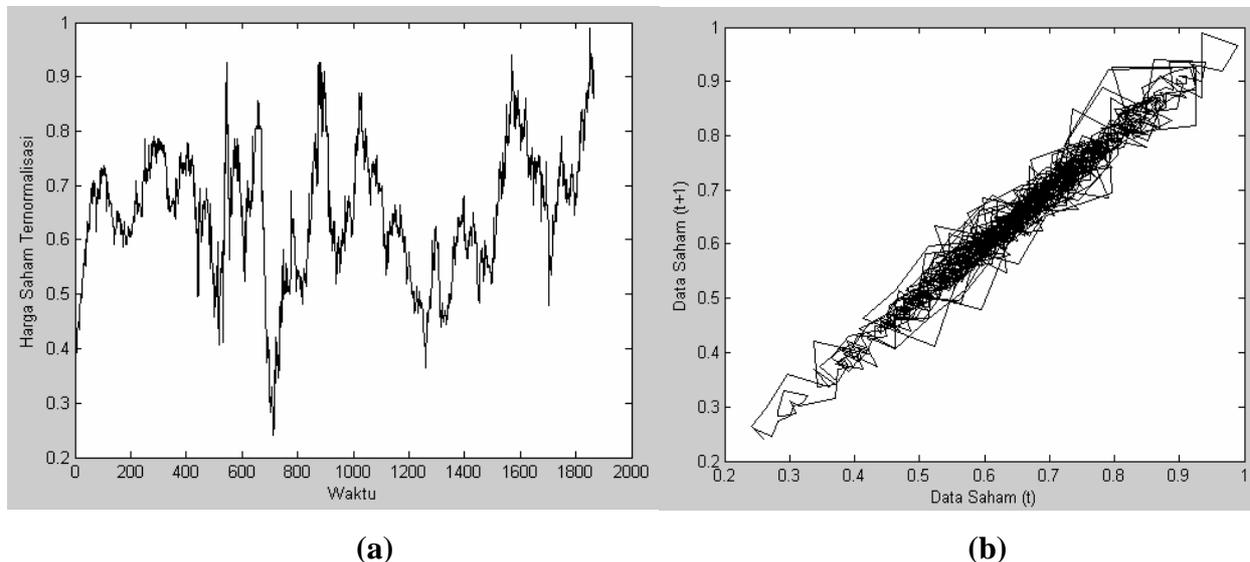

**(a)** **(b)**

**Figure 5**
PT TELKOM stock data (1993 –medio 2003) after normalized for prediction. (a) Time series data (b) Poincare map that will be percept by Neural Network

## 3. Implementative Case-Study: Stock Price of PT TELKOM

We try to make prediction on close stock price value of PT TELKOM for time series data from the early of 1995 to middle of 2003. Data we will predict is displayed in Figure 5. The time series data (Figure 5a) is transformed into the modified Poincare map henceforth we obtain fluctuation move pattern of the stock price data. This Poincare map is then percepted by using our artificial neural network. Topology of neural network we choose is 15-15-16-16-15-15-1 with transfer function for each layer is the sigmoid tangent function while the output layer is the pure linear function. Training rate we use is back-propagation using *updating* neuron weight factor linear gradient is $\alpha=0.05$ and momentum $\beta=0.5$. Figure 6 shows the training process from neural network we use.

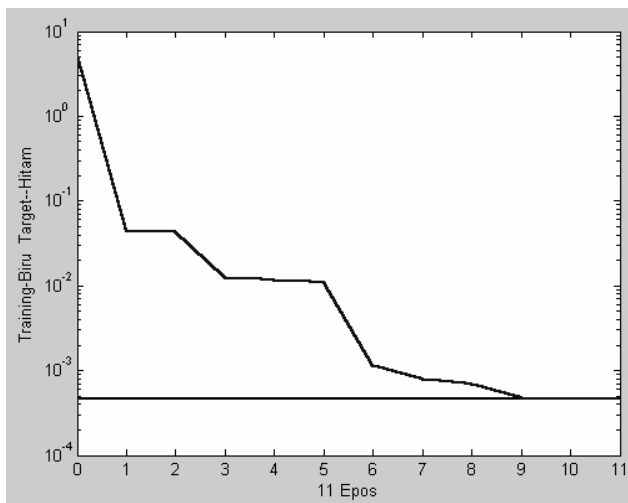

**Figure 6**
Learning process of neural network along perception of stock data Poincare map

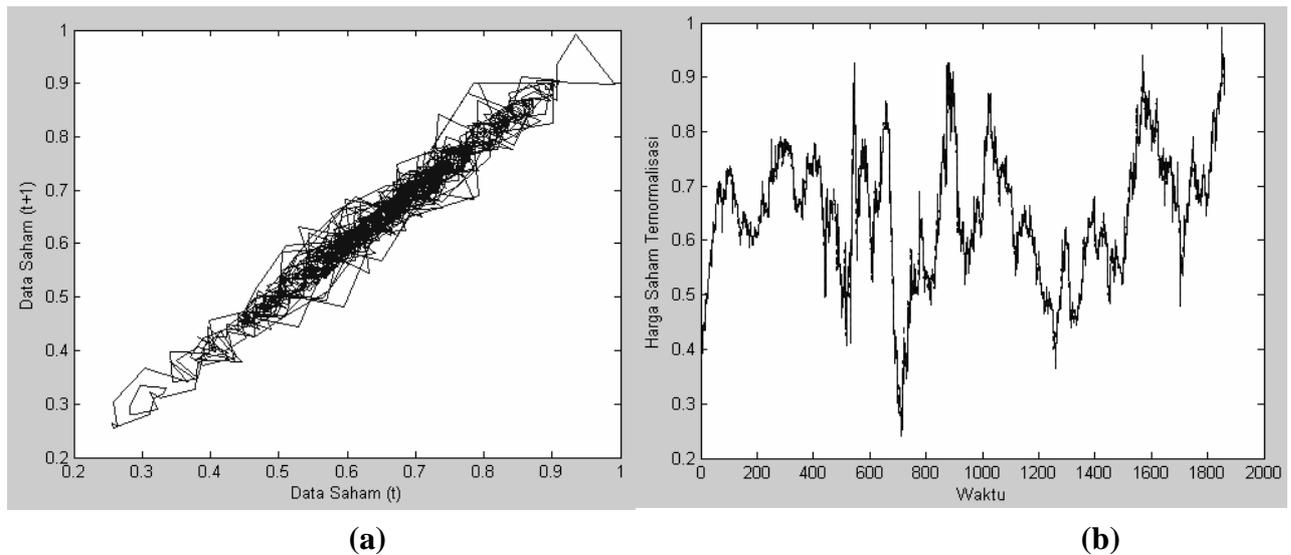

**(a)** **(b)**

**Figure 7**
Poincare map obtained (a), compared to 5b, and approximation of existing time series data after converted into time series

By using the perception process we will find out the result in the eleventh epoch. The result is the data in the form of the Poincare map as shown in Figure 7a. Approximation result of time series data is shown in Figure 7b. From here we can carry out prediction in further steps ahead. Prediction data is then adjusted with real time series data as seen in Figure 8. We can see obviously how the method of perception on the Poincare map can gain better and quicker result than the conventional time series part. It is clear since we don't percept time series data anymore, but from the pattern of the time series data.

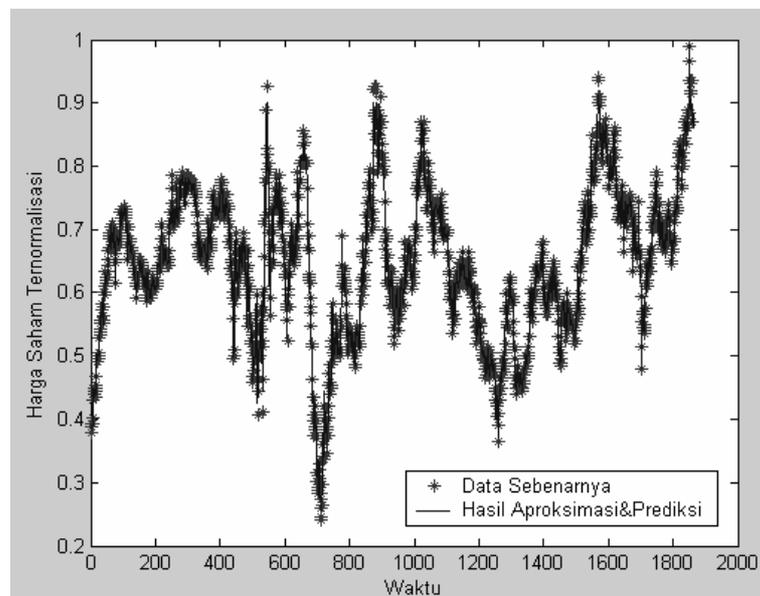

**Figure 8**
Approximation and prediction result altogether in time series picture

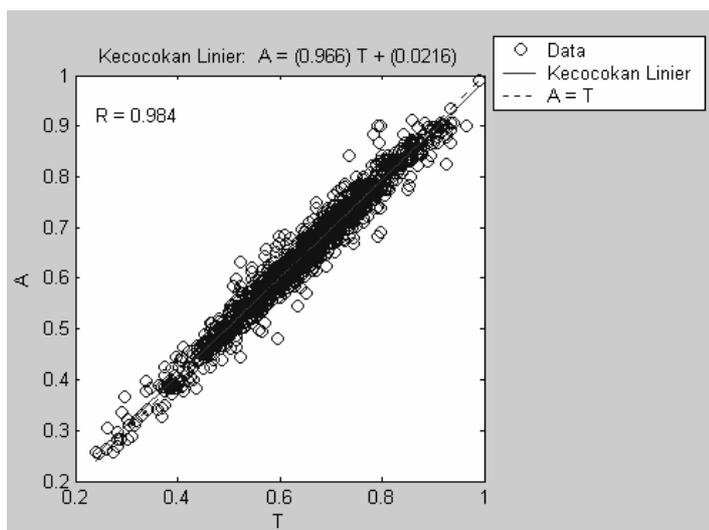

**Figure 9**
Linear suitability plot of the artificial neural network perception. It is shown that the approximation and prediction result is better with gradient approaches 1 ($\approx$0.966)

In order to ease the use of this prediction result, we also add an artificial neural network model predicting *range* of stock price value; here is believed so that this generic model can be used to more utilize predicted results. In this process, we use HIGH (highest daily price) and LOW (lowest daily price) data from time series data of fluctuating stock of PT TELKOM (1995-2003).

During the process, we change HIGH data into Poincare map and percept as in previous example. The difference approach in the following example is that we seek a difference of the highest price offering (HIGH) and the lowest (LOW) and this time series data difference then be analyzed using Poincare Map for further perception by using our artificial neural network.

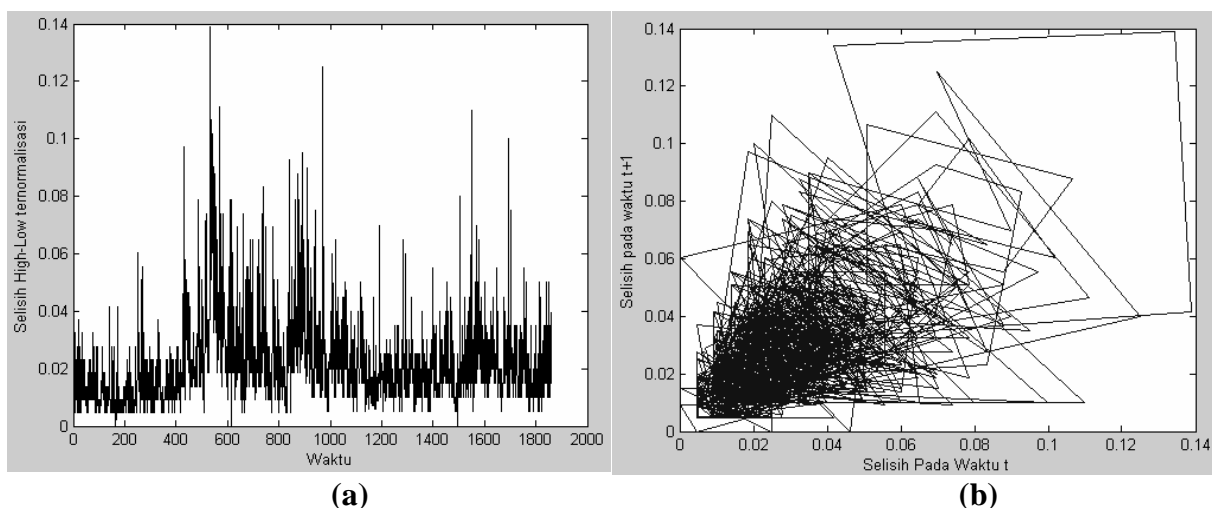

                **(a)**                              **(b)**

**Figure 10**
Difference data of the highest and the lowest offer of fluctuating stock price (a) and mapping result in Poincare map (b)

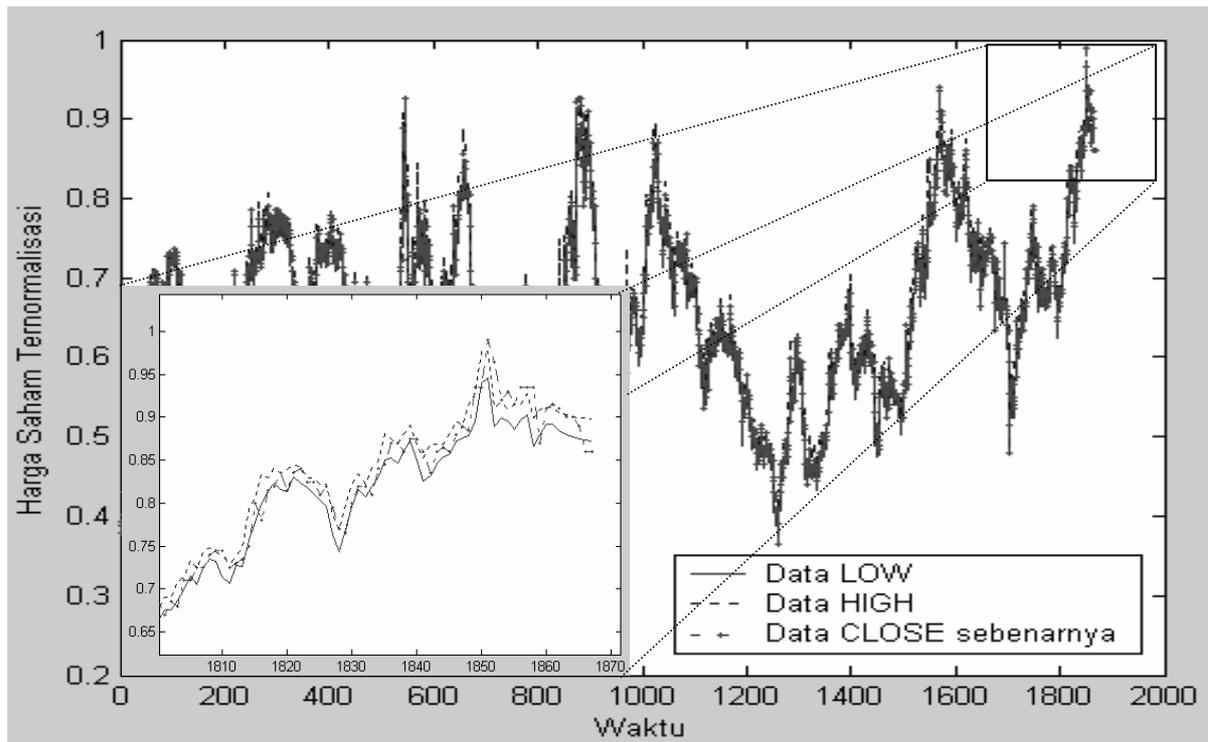

**Figure 11**
Approximation result the perception of the neural network of highest value and the
lowest value. The predicted is zoomed as shown in the inset; look out that the
*closing* data always lie in expected *range*

From this point, we start implementing second perception of the two neural networks that we construct particularly. Approximation and prediction result is in the form of *range* of value that possibly happens in the future. This is obviously seen in Figure 11, while we can also see the real *closing* data and that our prediction has already been fine with position of *closing* data lie between intervals of the approximated and predicted.

## 4. Conclusion

What we are trying to offer in the paper is a sharpening of perception by artificial neural network on financial time series data. We try to arrange algorithm that is not just predicting time series data in certain and usual way. The existed time series data is first transformed into a new Poincare map while then percepted by the neural network.

Poincare map is one of methodology approach of non-linear dynamical system, especially to a system with a large amount of time series data. Financial system in the form of the fluctuation of exchange value, ups-and-downs of financial data indexes and so forth are part of our social (economy) system whose huge number of data. Poincare map is pattern map of dynamic system trajectory, and a little modification of Poincare map has given us ability to "recognize" slightly movement of those big numbers of financial data.

Methodology offered in this paper is, of course still hypothetical, though the result obtained quite satisfying, examinations and comparative analysis even methodological combination with other approach surely will enrich prediction construction offered in this paper.

By this method, we will no longer design artificial neural network model to directly percept financial time series data as many has ever done these days in financial system. Neural network indirectly approximates those financial time series data. It is advantageous to use this methodology since its ability to gain more accurate perception of data - as pattern data (not a raw data) - and the better speed of computation relative to the previous methodology. We will no need to calculate more data, let the computer does it. We will no longer percept the data pattern, let the computer see and predict it. Let the computer be our loyal servant in facing non-linearity of the universe.

## Acknowledgement

The research reported by the paper is done on financial support from the Lembaga Pengembangan Fisika Indonesia. The authors thank Yohanis for the raw data from Indonesia stock exchange and to Deni Khanafiah for the assistance on processing the data. The authors also thank colleagues in BFI for the discussions on the rough version of the paper.